\documentclass{amsart}

\usepackage{epsfig}
\usepackage{graphicx}
\usepackage{amsmath,amssymb,amsfonts,latexsym}
\usepackage{graphicx}

\usepackage{amssymb}
\usepackage{enumerate, xspace}
\usepackage{dsfont}
\usepackage{afterpage}
\usepackage{changebar}
\usepackage{amsthm}
\usepackage{rotating}

\numberwithin{equation}{section}

\begin{document}

\title[Ion selectivity in ion channels]
{Generalized microscopic theory of ion selectivity in voltage-gated ion channels}
\author[Andrew Das Arulsamy]{Andrew Das Arulsamy} \email{sadwerdna@gmail.com}

\address{Institute of Mathematical Sciences, University of Malaya, 50603 Kuala Lumpur, Malaysia}
\address{Present Address: Condensed Matter Group, Division of Interdisciplinary Science, F-02-08 Ketumbar Hill, Jalan Ketumbar, 56100 Kuala-Lumpur, Malaysia}


\date{\today}

\begin{abstract}
Ion channels are specific proteins present in the membranes of living cells. They control the flow of specific ions through a cell, initiated by an ion channel's electrochemical gradient. In doing so, they control important physiological processes such as muscle contraction and neuronal connectivity, which cannot be properly activated if these channels go haywire, leading to life-threatening diseases and psychological disorders. Here, we will develop a generalized microscopic theory of ion selectivity applicable to KcsA, Na$_{\rm v}$Rh and Ca$_{\rm v}$ (L-type) ion channels. We unambiguously expose why and how a given ion-channel can be highly selective, and yet has a conductance of the order of one million ions per second, or higher. We will identify and prove the correct physico-biochemical mechanisms that are responsible for the high selectivity of a particular ion in a given ion channel. The above mechanisms consist of five conditions, which can be directly associated to these parameters---(i) dehydration energy, (ii) concentration of the ``correct'' ions (iii) Coulomb-van-der-Waals attraction, (iv) pore and ionic sizes, and indirectly to (v) the thermodynamic stability and (vi) the ``knock-on'' assisted permeation.
\end{abstract}

\maketitle

\section{Introduction}

Consciousness remains largely a philosophical entity, which has not been regarded as a serious scientific subject. The reason for this partly lies in the absence of a proper technical definition of what consciousness means (with respect to some physico-chemical notions). Its mysterious existence can be related to the real-time inputs from the five physical senses (hearing, sight, touch, smell and taste), or from within the brain itself (from stored memories), or both, such that the other extrasensory perceptions (for example, the sixth and higher senses) can be ignored for the time being. These extrasensory perceptions should not be confused with other types of physical senses such as the ability to detect magnetic and electric fields. For example, unlike humans, birds can biologically sense magnetic fields~\cite{bird} such that the \textit{birds do not need to solve any Maxwell equations}~\cite{bird2}, and similarly, certain fish can sense electric fields~\cite{fish}. However, it is due to our conscious mind that we are able to predict and measure the magnetic and electric fields by other means, without requiring to sense them biologically. Anyway, for humans, the above five physical senses can lead to the required excitory and inhibitory neuron-firing such that the firing of these neurons can activate the proper Neural Correlate(s) Consciousness (NCC), which can be quantitatively different from other animals and insects~\cite{crick,crick2}. The quantitative differences here can be due to the types of neurons and physical senses, and the number of neurons. In particular, the evolution of human SRGAP2 gene may be responsible for the molecular and cellular structure of a neuron in humans, giving rise to different types of neurons in humans compared to non-human mammals~\cite{den,char}.  

We anticipate that a scientific definition for consciousness based on the physico-chemical notions can be developed if we could first understand the underlying microscopic mechanisms of ion channels in neurons where neuronal signaling are the very basis for the existence of a conscious mind---\textit{what is stronger than fate? if we think of an expedient $($to avert it$)$, it will itself be with us before the thought}~\cite{kural}. For example, certain types of ion channels in brains (in neurons) are responsible for complicated neuronal signaling, which in turn gives rise to the thinking process. Therefore, a proper scientific definition can spearhead the ``consciousness'' as a valid scientific endeavor. To do that, we need to first study the ion channels, and their ability to transfer ``physico-chemical'' information \textit{via} the movement of ions through these ion channels that give rise to neuronal signaling. Of course, we will ignore the effect of drugs or nanoparticles on ion channels, for example, drugs~\cite{sel} and nanoparticles~\cite{alice} can enhance and/or suppress certain ion channels, which can lead to permanent damages to neurons.  

Generally, ion channels are transmembrane proteins consist of pore forming subunits and accessory subunits, which are thought to be encoded in no less than 340 human genes~\cite{ash}. Each subunit is a collection of large number of amino acids (of the order of 10$^2$ to 10$^3$). Unlike ion transporters (that require two different ion species (with opposite charges) moving in the opposite direction) and ion pumps (that require energy supply from ATP (Adenosine-5$'$-triphosphate) hydrolysis), the voltage-gated ion channels are activated by the changes in electrical potential (stimulus) in the vicinity of a cell~\cite{ash,kandel}. For example, the positive charge concentration within a cell is different from the positive charge content outside the cell, which is the source for an electrochemical gradient. 

Moreover, the voltage-gated ion channels contain selectivity filters formed by $\alpha$-helices (with residues containing negatively charged carbonyl and/or carboxyl group oxygens)~\cite{hil}. Therefore, each of these ion channels has evolved to only allow one type of ions to permeate through the selectivity filter with maximum conductance. For example, KcsA and K$_{\rm v}$1.2 ion channels efficiently select and permeate K$^+$ maximally, which also systematically exclude Na$^+$ and other monovalent (Rb$^+$, Cs$^+$) and divalent cations (Sr$^{2+}$, Ca$^{2+}$, Mg$^{2+}$, Ba$^{2+}$, Cd$^{2+}$, Co$^{2+}$)~\cite{kandel,hil}. Anions are excluded due to Coulomb repulsion between the negatively charged carbonyl (or carboxyl) group oxygen and a given anion~\cite{hil,madh}. Of course there are other factors involved that keep away the anions from entering the cation-favoring channels, but the Coulomb repulsion has made sure of that (non-entry for anions).  

Here, we will only consider the cation-favoring voltage-gated ion channels, and show that the selection criteria of these particular ion channels have got everything to do with the physical properties of electrons, atoms, ions, molecules, and their interactions, namely, the atomic and molecular energy-level spacings, the electron-electron and electron-ion Coulomb interactions and the stronger van der Waals (vdW) attraction. Moreover, the criteria also include the selectivity pore size and ionic size. Our analyses presented here exploit the KcsA (from the soil bacteria \textit{Streptomyces lividans}~\cite{mac,way}), Na$_{\rm v}$Ab (from the bacterium \textit{Arcobacter butzleri}~\cite{jian,jian2}), Na$_{\rm v}$Rh (from the NaChBac \textit{alphaproteobacterium HIMB114})~\cite{xu}) and the L(long lasting activation)-type Ca$_{\rm v}$~\cite{hess,will} ion channels. We anticipate that the generalized mechanism developed and discussed herewith can be extended to understand other types of ion channels, namely, the ion pumps and ion transporters, as well as the inwardly-rectifying K$^+$ (K$_{\rm ir}$) channels. The ionic mass effect (for example, $M_{\rm K^+} > M_{\rm Na^+}$) can be neglected because the electronic (between electrons) and ionic (between ions or atoms from different residues or molecules) interactions are much larger than the effects originating from different ionic masses.       

Our objective here is to develop and justify a microscopic theory for ion selectivity in ion channels that can be used to determine the precise physico-biochemical mechanisms responsible for the high specificity of ion channels. For example, we need to understand why and how the KcsA, Na$_{\rm v}$Rh, and the L-type Ca$_{\rm v}$ ion channels can select and permeate the respective K$^+$, Na$^+$ and Ca$^{2+}$ ions maximally. This means that, we need to uncover the electron-electron and electron-ion interactions between atoms from different residues or molecules starting from the first principles (without any guessed atomic orbitals or wavefunctions and adjustable parameters). Our focus and aims here are to first identify the types of physical interactions (down to the electronic level) that are responsible for the ion channels selectivity. In particular, the physical properties include the effect of atomic polarization, the electron-electron (Coulomb type) and the electron-ion (van der Waals and Coulomb types) interactions. 

We now provide the relevant details as to the reasons why we need to revisit the ion selectivity properly such that it goes beyond the thermodynamical and kinetic approaches. First warning: this paper is on ion selectivity of ion channels when the gates are in an open configuration, and therefore, we will not discuss the gating mechanism(s) that is(are) responsible for the opening and closing of the voltage-gated ion channels. Second warning: gating mechanisms may not be completely independent of ion selectivity as reported by Lockless et al.~\cite{lock} in the wild-type KcsA such that the selectivity pore assumes a collapsed configuration in the absence of K$^+$ ions. This is a valid point both logically and scientifically because the gating processes can be controlled by the concentration of the correct ions in the vicinity of a cell. Fortunately, this concentration-dependent gating will not invalidate the theory of ion selectivity developed here because the said theory by definition becomes inapplicable when the channels are closed. This implies that the theory has been designed in such a way that we can ignore the closing mechanism---it does not matter whether the closing is due to the absence of the ``correct'' ions, or not.
        
Important advances were made on ion selectivity after the discovery of pore-forming protein crystal structures, namely, the KcsA (K$^+$), Na$_{\rm v}$Ab (Na$^+$) and Na$_{\rm v}$Rh (Na$^+$) ion channels by Doyle et al.~\cite{mac}, Payandeh et al.~\cite{jian} and Zhang et al.~\cite{xu}, respectively. The minimum pore size in a KcsA channel is between 3~\AA~ and 4~\AA~\cite{clay,kin}, which allows a partially hydrated K$^+$ ion to permeate. Partial hydration here means that a single K$^+$ is attached to at most two water molecules such that a single K$^+$ is sandwiched between two water molecules. The partially hydrated H$_2$O$---$K$^+$$---$OH$_2$ complex enters the narrowest pore partly coordinated by the carbonyl group oxygens (from the selectivity pore). The dashed lines ``$---$'' denotes the Coulomb-vdW attraction. If these lines are drawn between a negatively charged oxygen and a positively charged hydrogen atoms, then it is known as hydrogen bonds (usually occurs between water molecules).  

The above mechanism has been verified to be true compared to the experimental crystal structure data for KcsA, which contains the negatively charged backbone carbonyl group oxygens that form the narrowest pore~\cite{mac}, which also nicely fit a partially hydrated K$^+$~\cite{clay}. A partially hydrated Na$^+$ is too small to be stably coordinated by these carbonyl oxygens~\cite{clay}. In other words, a Na$^+$ ion does not fit in the selectivity pore ``nicely'' with a better thermodynamic stability compared to a K$^+$ ion. 

Theoretical validation in favor of K$^+$ (over Na$^+$) using the molecular dynamical/quantum mechanical methods were reported in Refs~\cite{mul,mul2,nos,roux}. In particular, thermodynamically (by means of free-energy and dehydration-energy calculations), a partially hydrated K$^+$ ion coordinated in the selectivity pore has a lower free energy (more stable) compared to a partially hydrated Na$^+$, in agreement with Refs.~\cite{beza,hille}. In these calculations~\cite{nos}, one need not enforce some fixed positions for the carbonyl oxygens (that are flexible to some extent), and the dehydration energy for a hydrated Na$^+$ is larger than that of a hydrated K$^+$. Therefore, both free-energy and dehydration energy~\cite{gann} calculations favor K$^+$ over Na$^+$ for permeation through a KcsA ion channel. Moreover, one should be aware here that the influence coming from the number of coordinating ligands (within the filter) on K$^+$ selectivity has been proposed in Ref.~\cite{jaya}, but this is just an alternative way to reinforce the thermodynamical effect highlighted earlier. For example, the optimized number of coordinating ligands are indeed required for optimum permeation with minimal energy barrier (low free energy) along the correct ions conducting pathway.    

However, the thermodynamic stability discussed above does not tell us ``enough'' about the selectivity mechanism, for example, the incorrect ion (Na$^+$), after it enters the pore, is not ejected out of the pore because the Na$^+$ ion is now found to be thermodynamically unstable (compared to a K$^+$ ion) within the KcsA selectivity pore. In fact, there is no such thing as ``ejection mechanism'', and it is not responsible for ion selectivity. The above negation of the ejection mechanism is true regardless whether a given selectivity pore contains a single binding site or multiple binding sites. Interestingly, there are also other binding sites within the KcsA selectivity filter, which thermodynamically favor Na$^+$ over K$^+$~\cite{thomp}. It so happens that in a KcsA selectivity pore, one has four binding sites for K$^+$, giving rise to another hypothesis---the snug-fit mechanism~\cite{ney,ney2,mac,cri}. Similar to the thermodynamical approach stated above, the snug-fit hypothesis does tell us the mechanism for K$^+$ permeation through the KcsA selectivity pore that contains four binding sites, but it still lacks the ability to explain why and how a KcsA ion channel selects K$^+$ over Na$^+$ (or any other monovalent and divalent ions). The second reason why thermodynamical stability is not a direct criterion for ion selectivity (at least for KcsA) is because there are alternative binding sites, which are energetically favorable to Na$^+$ within the selectivity filter (with lower free energy compared to K$^+$)~\cite{thomp}.

Therefore, it is worth noting that the question of ion selectivity remains elusive since the first measurements of electrical signaling in nerve cells by Hodgkin and Huxley in the early 1950s~\cite{hod,varma,dixit}. In particular, the permeation mechanism in a KcsA ion channel has been explained in Refs.~\cite{hod2,kuyu,kuyu2,roux2} by means of the balancing forces among the positively charged ion-ion Coulomb repulsion (in the presence of coordinating carbonyl oxygens), which in turn leads to the ``knock-on'' assisted multi-ion permeation mechanism~\cite{hod2,kuyu,kuyu2,roux2}. The knock-on hypothesis, which gives rise to the kinetic approach indeed controls the magnitude of ionic conductance in a KcsA channel, but it is not directly related to ion selectivity due to the nonexistent of any ejection mechanism as pointed out earlier. This means that, the ``incorrect'' ions are not ejected out of the filter once they (the incorrect ions) have entered the selectivity pore, regardless whether the incorrect ions are thermodynamically stable or not within the selectivity filter. This argument is also applicable for Na$_{\rm v}$Rh~\cite{xu} and the L-type Ca$_{\rm v}$~\cite{hess,will} ion channels (we will return to these channels at a later stage). Apparently, our motivation here is to establish the microscopic mechanisms that are responsible for ion selectivity in voltage-gated cation-favoring ion channels, beyond the thermodynamical-stability and ion-permeation approaches. We now proceed to introducing the relevant theory required to develop a generalized theory of ion selectivity in voltage-gated ion channels in the open configuration (when the gates are opened).

\section{Theoretical details}

A given biosystem that consists of ions and molecules should strictly obey the rules of quantum mechanics, which is observable at the molecular level. These quantum-mechanics obeying ions and molecules are the ones that gave us the license to invoke IET in biological systems. This also means that, a given biological system is not merely a mathematico-statistical mechanical system, though the biosystems behavior can be captured via the computational models relying entirely on mathematics and statistical mechanics. To put it another way, the biosystems microscopically obey some underlying physico-chemical phenomena, which are governed by the quantum mechanical notions, which will be unambiguously exposed here. 

A formal proof in favor of the above argument is given in Ref.~\cite{pccp} (see the analyses on dielectric properties of water depicted in Fig.3 in Ref.~\cite{pccp}). The said analyses show that the exponential behavior can be obtained from IET and quantum mechanics~\cite{pccp} down to an electronic level, while a macroscopic power-law behavior~\cite{uemat} was obtained from the mathematico-computational model. Another proof in this respect is given in Ref.~\cite{pssb} where the resistivity versus temperature curve in the ferromagnetic regime of a diluted magnetic semiconductor is proven to be exponentially driven, even though the said curve can also be computationally fitted with a power-law (see the discussion on Fig.5 in Ref.~\cite{pssb}). The second reason to use IET here is due to the fact that some of the quantum effects arising from the physical systems have been well developed in condensed matter physics within IET. The final reason is due to a hypothesis put forth by Baskaran---\textit{all basic condensed matter phenomena and notions mirror in biology}~\cite{bask}.      

\subsection{Ionization energy theory}

The ionization energy theory (IET) starts from the IET-Schr$\ddot{\rm o}$dinger equation~\cite{pra}, 
\begin {eqnarray}
&&i\hbar\frac{\partial \Psi(\textbf{r},t)}{\partial t} = \bigg[-\frac{\hbar^2}{2m}\nabla^2 + V_{\rm IET}\bigg]\Psi(\textbf{r},t) = H_{\rm IET}\Psi(\textbf{r},t) = (E_0 \pm \xi)\Psi(\textbf{r},t). \label{eq:1}
\end {eqnarray}  
Here, $\Psi(\textbf{r},t)$ is the usual time-dependent many-body wave function, $\hbar$ is the Planck constant divided by $2\pi$ and $m$ is the mass of electron. Contrary to the standard Schr$\ddot{\rm o}$dinger equation, which directly deals with the energy eigenvalue, $E$, the IET-Schr$\ddot{\rm o}$dinger equation on the other hand requires an eigenvalue that reads $E_0 \pm \xi$ (see Eq.~(\ref{eq:1})), which has been proven to be equal to the standard eigenvalue, $E$. Here, $E_0$ is defined to be the energy levels at zero temperature, and also without any external disturbances. Similar to $E$, $E \pm \xi$ also represent the real (true and unique) energy levels for a particular quantum system. 

However, $E_0 \pm \xi$ carries an additional microscopic information such that $E_0$ is a constant energy eigenvalue, and the additional information comes from $\xi$, which is defined to be the ionization energy (or the energy-level spacing). The sign, ``$\pm$'' refers to electrons and holes, respectively. However, we need to make use of the ionization energy approximation to determine $\xi$. The ionization energy approximation reads $H_{\rm IET}\Psi(\textbf{r},t) = (E_0 \pm \xi^{\rm quantum}_{\rm matter})\Psi(\textbf{r},t) \propto (E_0 \pm \xi^{\rm constituent}_{\rm atom})\Psi(\textbf{r},t)$, which has been proven in Ref.~\cite{pra} such that the proof has been associated to the Shankar screened Coulomb potential~\cite{rmp} and the ionization energy based Fermi-Dirac statistics~\cite{physc,pla}. This approximation can also be written in the following form
\begin {eqnarray}
&&\xi^{\rm quantum}_{\rm matter} \propto \xi^{\rm constituent}_{\rm atom}. \label{eq:2}
\end {eqnarray}  
As introduced earlier, $\xi^{\rm quantum}_{\rm matter}$ is the real and unique energy level spacing of a particular quantum system. This real energy level spacing is the energy cost that needs to be overcome by an electron that tries to occupy another energy level. In atoms and ions, $\xi$ is known as the atomic energy level spacing, and for molecules, $\xi$ obviously refers to the molecular energy level spacing. Therefore, using Eq.~(\ref{eq:2}), one can predict the changes that may occur in $\xi^{\rm quantum}_{\rm matter}$ by calculating the values for $\xi^{\rm constituent}_{\rm atom}$. The experimental values for $\xi^{\rm constituent}_{\rm atom}$ are available in atomic spectra and other databases~\cite{web,web2}. The ionization energy approximation given in Eq.~(\ref{eq:2}) becomes exact for atoms or ions because $\xi^{\rm quantum}_{\rm matter} = \xi^{\rm atom}_{\rm ion}$.  

The data for $\xi^{\rm constituent}_{\rm atom}$ can be obtained from Refs.~\cite{web,web2}, and therefore, one can readily exploit the (ionization energy) approximation to determine the required physico-chemical properties, which in turn will enable us to understand why and how a particular biochemical property changes with changing interaction strengths. In real systems however, interactions between atoms or ions or molecules do not necessarily involve a single valence electron, consequently, for such interactions that involve more than one electron (from an atom), one needs to average the atomic ionization energies for atoms that donate (or share) more than one electron. The said averaging follows 
\begin {eqnarray}
&&\xi^{\rm constituent}_{\rm atoms} = \sum_j\sum_i^z \frac{1}{z}\xi_{j,i}(\texttt{X}^{i+}_{j}). \label{eq:3}   
\end {eqnarray}
Here, the subscript $j$ identifies the types of chemical elements ($\texttt{X}_j$) that exist in a particular molecule. The other subscript, $i = 1, 2, \cdots, z$, counts the number of valence electrons originating from a particular chemical element. For example, the average ionization energy for Cd$^{2+}$ is given by  
\begin {eqnarray}
&&\xi_{\rm Cd^{2+}} = \frac{1}{2}\sum_i^2 \xi_{{\rm Cd}^{i+}} = 1250~ {\rm kJmol^{-1}}, \label{eq:4}   
\end {eqnarray}
which means that one requires an energy proportional to 1250~kJmol$^{-1}$ to excite one of the two valence electrons from an atomic Cd. In other words, each of these two electrons needs an average energy proportional to 1250~kJmol$^{-1}$ to be excited or polarized. All the ionization energies prior to averaging were obtained from Refs.~\cite{web,web2}. Alternatively, the required ionization energies for isolated atoms and ions can be calculated from the density functional theory using some guessed wavefunctions and variationally adjustable parameters~\cite{dft}. We alert the readers here that the ionization energy theory belongs to the class of renormalization group theory~\cite{pra,aop}, which can be related exactly to the Shankar renormalization group method within the screened Coulomb potential~\cite{rmp,shank2,shank3}.

\subsection{Generalized van der Waals interaction}

The standard van der Waals interaction~\cite{grif} 
\begin {eqnarray}
&&V^{\rm std}_{\rm Waals}(R) = -\frac{\hbar}{8m^2\omega_0^3}\bigg(\frac{e^2}{2\pi\epsilon_0}\bigg)^2\frac{1}{R^6}. \label{eq:5}
\end {eqnarray}  
Whereas, the renormalized vdW attraction has been derived elsewhere~\cite{pccp2}, 
\begin {eqnarray}
&&\tilde{V}_{\rm Waals}(R,\xi) = \bigg\{-\frac{\hbar}{8m^2\omega_0^3}\bigg(\frac{e^2}{2\pi\epsilon_0}\bigg)^2\frac{1}{R^6}\bigg\}\exp\bigg[-\frac{3}{2}\lambda\xi\bigg]. \label{eq:6}
\end {eqnarray}  
Subsequently, a generalized vdW attraction has been formally proven to be in the form of~\cite{pccp2}
\begin {eqnarray}
\tilde{V}'_{\rm Waals}(\xi) = V^{\rm e-ion}_{\rm Coulomb} + \frac{1}{2}\tilde{V}_{\rm Waals}(\xi), \label{eq:7}
\end {eqnarray}  
where $\tilde{V}_{\rm Waals}(\xi)$ is not given by Eq.~(\ref{eq:6}), instead, one should use the renormalized $R$-independent vdW attraction formula,
\begin {eqnarray}
\tilde{V}_{\rm Waals}(\xi) = \hbar\omega_0\bigg(\frac{1}{\sqrt{2}} - 1\bigg)\exp{\bigg[\frac{1}{2}\lambda\xi\bigg]}, \label{eq:8}
\end {eqnarray}  
to be substituted for $\tilde{V}_{\rm Waals}(\xi)$ in Eq.~(\ref{eq:7}), or alternatively, $\tilde{V}_{\rm Waals}(\xi)$ can be substituted with  
\begin {eqnarray}
V_{\rm Waals}(R) = \bigg[\frac{\hbar^2e^2}{2m\pi\epsilon_0}\bigg]^{\frac{1}{2}}\bigg(\frac{1}{\sqrt{2}} - 1\bigg)\frac{1}{R^{3/2}}, \label{eq:9}
\end {eqnarray}  
where $V_{\rm Waals}(R) = \tilde{V}_{\rm Waals}(\xi)$~\cite{pccp2}. Furthermore, the first term on the right-hand side of Eq.~(\ref{eq:7}) is given by
\begin {eqnarray}
V^{\rm e-ion}_{\rm Coulomb} = \frac{1}{4\pi\epsilon_0}\bigg[\frac{(-e)_{\rm electron}(+e)_{\rm ion}}{R}\bigg]. \label{eq:10}
\end {eqnarray}  
Importantly, Eq.~(\ref{eq:7}) gives a stronger vdW attraction, stronger than the standard vdW interaction (Eq.~(\ref{eq:6})). Here, $R$ is the separation between two nuclei, $\hbar\omega_0$ is the averaged ground state energy of a system that consists of an ion interacting with carbonyl oxygens in the absence of external disturbances (for $T = 0$K). Here, $e$ is the electron charge, $\lambda = (12\pi\epsilon_0/e^2)a_{\rm B}$, $a_{\rm B}$ denotes the Bohr radius of an atomic hydrogen and $\epsilon_0$ is the permittivity of free space. A formal derivation for Eqs.~(\ref{eq:6}) and~(\ref{eq:7}) are given in Ref.~\cite{pccp2}, which includes the physical insights on why and how a stronger attraction can be invoked from Eq.~(\ref{eq:7}). Note here that the relevant equations in our case are Eqs.~(\ref{eq:7}) and~(\ref{eq:8}) because (as explained earlier) we need equations that are $\xi$-dependent. We now have one more theoretical issue to be settled before digging deep into the mechanisms of ion selectivity in cation-favoring ion channels.  

\subsection{Physical properties of ions}

Here, we provide some established properties of ions, which are required for our analyses later. This is followed by the proper interpretations on some of these physical properties. Table~\ref{Table:I} lists the chemical elements, their averaged ionization energies, valence states and ionic sizes that are relevant to this work. One important fact to be noted here is that a cation with the largest averaged ionization energy needs large energy to excite or polarize its valence electron. Alternatively, this large ionization-energy cation can attract the electrons (with a stronger attractive force) from a neighboring ion, provided that this neighbor has a lower averaged ionization energy. In all of our cases here, the neighbor remains the same---carbonyl/carboxyl group oxygens, and therefore, we just need to determine the relative ionization energy values solely for the cations. Cations with a valence state, 1+ do not require averaging (see Eq.~(\ref{eq:4})). In contrast, for all other valence states larger than 1+, one has to average a cation's individual ionization energies (see Eq.~(\ref{eq:3})). Individual ionization energies here refer to the first, second, $\cdots$ ionization energies, such that the first ionization energy equals the energy required to remove the first outer (valence) electron to infinity, and so on. From here onwards, we will drop the word ``averaged'' when we discuss the ionization energy values, for it is obvious from Eqs.~(\ref{eq:3}) and~(\ref{eq:4}). 

\begin{table}[ht]
\caption{Averaged atomic ionization energies ($\xi$) are listed below for individual ions. The chemical elements are ordered with increasing atomic number $Z$, which include their ionic sizes for the cations. The experimental ionization energy values prior to averaging were obtained from Refs.~\cite{web,web2}, and the averaging follows Eq.~(\ref{eq:3}). We use the unit kJmol$^{-1}$ instead of eV$\cdot$atom$^{-1}$ for convenience.} 
\begin{tabular}{l c c c c} 
\hline\hline 
\multicolumn{1}{l}{Element}         &    Atomic number  &  Valence     & Ionic size                & $\xi$   \\  
\multicolumn{1}{l}{}                &   $Z$             &  state       & (diameter, \AA)  &(kJmol$^{-1}$)\\  
\hline 

H                                   &  1   					    &  1+          & 0.5   & 1312 \\ 
Li                                  &  3	   	  			  &  1+          & 1.2   & 520  \\ 
C                                   &  6                &  4+          & ---   & 3571 \\
O                                   &  8	   	  			  &  1+          & ---   & 1314 \\ 
O                                   &  8	   	  			  &  2+          & ---   & 2351 \\ 
O                                   &  8	   	  			  &  4+          & ---   & 4368 \\ 
Na                                  &  11	   	  			  &  1+          & 1.9   & 496  \\ 
Mg                                  &  12	   	  			  &  2+          & 1.3   & 1094  \\
K                                   &  19	   	  			  &  1+          & 2.67  & 419  \\
Ca                                  &  20	   	  			  &  2+          & 1.98  & 868  \\
Co                                  &  27	   	  			  &  2+          & 1.44  & 2408  \\
Rb                                  &  37	   	  			  &  1+          & 2.96  & 403  \\ 
Sr                                  &  38	   	  			  &  2+          & 2.26  & 807  \\
Cd                                  &  48	   	  			  &  2+          & 1.94  & 1250  \\ 
Cs                                  &  55	   	  			  &  1+          & 3.38  & 376  \\ 
Ba                                  &  56	   	  			  &  2+          & 2.7   & 734  \\

\hline  
\end{tabular}
\label{Table:I} 
\end{table}
The second point you should note from Table~\ref{Table:I} is that not all 1+ cations interact equally, with an equal magnitude of Coulomb force with a given anion (for example, with a carbonyl oxygen). In particular, K$^+$ and Na$^+$ are both effectively positively charged, 1+~\cite{kandel}, however, screening effect due to electron-electron and electron-ion interactions give rise to different electron affinities for these cations to attract electrons from a neighboring anion or atom. These anions or atoms may come from a different molecule. Using IET we can readily deduce that a Na$^+$ (1.9~\AA) can attract an electron with a much stronger Coulomb force compared to a K$^+$ (2.67~\AA) because $\xi_{\rm K^+}$ (419~kJmol$^{-1}$) $<$ $\xi_{\rm Na^+}$ (496~kJmol$^{-1}$) (see Table~\ref{Table:I}). Here, Na$^+$ being smaller than K$^+$ has got nothing to do with their ability to attract electrons (or their electron affinity). In other words, any association that may exists between ionic size and electron affinity is just a ``lucky coincidence''. This means that (microscopically), a cation with large $\xi$ can attract electrons with a stronger Coulomb force compared to a cation with a lower $\xi$. The readers are referred to Refs.~\cite{pra,aop} for the formal proofs on IET and its approximation. In these Refs.~\cite{pra,aop} we also provide detailed explanations as to why and how the above interactions come to play and influence the electron affinity of an ion or atom. 

The importance of ionic size is not obvious from the above examples, however, it becomes clear when we revisit the reason why a Na$^+$ does not fit properly (or nicely) compared to a K$^+$ in a KcsA selectivity filter (as discussed earlier). In particular, the separation between a cation (Na$^+$ or K$^+$) and a carbonyl oxygen, $R$ remains the same for both Na$^+$ (1.9~\AA) and K$^+$ (2.67~\AA) because $R$ actually measures the distance between two nuclei centers ($R = \big|\textbf{R}^{\rm ion}_{\rm K^+,Na^+} - \textbf{R}^{\rm carbonyl}_{\rm O}\big|$). However, the outer most electron of a K$^+$ ion is in fact closer to a carbonyl oxygen (compared to Na$^+$) because the valence electron from a K$^+$ forms a larger Fermi surface, which also interacts weakly such that a K$^+$ ion can be easily stabilized (compared to a smaller Fermi surface of a Na$^+$ ion) by the valence electron from a carbonyl oxygen. In other words, the effective electron-electron distance ($\big|\textbf{r}^{\rm K^+}_1 - \textbf{r}^{\rm O}_2\big|$) between a K$^+$ ion and a carbonyl oxygen is smaller than the electron-electron distance between a Na$^+$ ion and a carbonyl oxygen. This means that, $\big|\textbf{r}^{\rm Na^+}_1 - \textbf{r}^{\rm O}_2\big| > \big|\textbf{r}^{\rm K^+}_1 - \textbf{r}^{\rm O}_2\big|$ exists due to a larger Fermi surface of a K$^+$ ion compared to a Na$^+$ ion where $\big|\textbf{R}^{\rm ion}_{\rm Na^+} - \textbf{R}^{\rm carbonyl}_{\rm O}\big| = \big|\textbf{R}^{\rm ion}_{\rm K^+} - \textbf{R}^{\rm carbonyl}_{\rm O}\big|$. It is due to this inequality, $\big|\textbf{r}^{\rm Na^+}_1 - \textbf{r}^{\rm O}_2\big| > \big|\textbf{r}^{\rm K^+}_1 - \textbf{r}^{\rm O}_2\big|$ and a strong interaction (between a Na$^+$ and a carbonyl oxygen) that one can logically show why a Na$^+$ ion does not properly fit into a KcsA selectivity pore. Here, a Na$^+$ ion is guaranteed to interact more strongly (compared to a K$^+$) with an electron from a carbonyl oxygen because $\xi_{\rm Na^+}$ (496~kJmol$^{-1}$) $>$ $\xi_{\rm K^+}$ (419~kJmol$^{-1}$) and $\big|\textbf{R}_{\rm Na^+} - \textbf{r}^{\rm O}_2\big| = \big|\textbf{R}_{\rm K^+} - \textbf{r}^{\rm O}_2\big|$. We are basically done exposing the relevant theoretical details required to proceed to the next stage of developing the ion selectivity mechanisms.     

\section{Generalized mechanisms for ion selectivity}

One of the most important measurable observables to characterize ion channels is their conductance. Conductance is measured with respect to time and concentration of various ion species, and therefore, we need to make contact with it, at least indirectly. This means that, we will not attempt to reproduce the conductance, which has been done by many researchers using macroscopic models~\cite{hil,hod,bastug}. As we have stressed in the introduction, our intention here is to come up with a microscopic theory that can capture the ion selectivity such that our theory is also required to explain why and how the intrinsic conductance for a given ion and ion channel changes in the presence of different ions. For example, we should be able to explain (down to an electronic level) (I) the correctness of the Eisenman sequences~\cite{hess,will,eisen} and (II) why and how the Na$^{+}$ current ($I_{\rm Na}$) is more effectively blocked by Cd$^{2+}$ (1.94~\AA) ions compared to Ca$^{2+}$ (1.98~\AA) in a Na$_{\rm v}$Rh ion channel~\cite{xu} (see Fig.2f in Ref.~\cite{xu}). 

The macroscopic models for ion channels developed thus far do take the intrinsic conductance of an ion in a given ion channel into account, but it is treated as a constant~\cite{hil,hod,bastug,madh2}. Subsequently, this constant is adjusted (by means of some guessed functions or hypothesis) whenever one changes the type of ions and/or the type of ion channels~\cite{hil,hod,bastug,madh2} to fit the experimental conductance data. The reason why the intrinsic conductances or other intrinsic parameters are treated as constants are explained in the following four examples. We call certain parameters as ``intrinsic constants'' because they can be related to some microscopic quantum mechanical notions, namely, the electronic wavefunctions or electronic energy levels, for a given system and for a set of conditions.\\
\textit{Example 1}: The standard Hodgkin-Huxley model behaves according to~\cite{adam} 
\begin {eqnarray}
&&\frac{{\rm d}V_{\rm m}}{{\rm d}t} = -\frac{1}{C_{\rm m}}\big[(V_{\rm m} - E_{\rm L})g_{\rm L} + (V_{\rm m} - E_{\rm K})g_{\rm K} + (V_{\rm m} - E_{\rm Na})g_{\rm Na} - I_{\rm injection}\big], \label{eq:11}
\end {eqnarray}  
where $I_{\rm injection}$ is a constant current, specific to an experiment, $C_{\rm m}$ and $V_{\rm m}$ denote the capacitance and the potential across a cell membrane, respectively, $g_{\rm L}$ is the conductance due to leakage, while $g_{\rm K}$ and $g_{\rm Na}$ are the ionic conductances (for K$^+$ and Na$^+$, respectively). Moreover, the reversal potentials due to leakage, K$^+$ and Na$^+$ are respectively denoted by $E_{\rm L}$, $E_{\rm K}$ and $E_{\rm Na}$. The point here is, $g_{\rm K} = \bar{g}_{\rm K}A^{\alpha}B^{\beta}$ and $g_{\rm Na} = \bar{g}_{\rm Na}A^{\alpha}B^{\beta}$ in which, $\bar{g}_{\rm K}$ and $\bar{g}_{\rm Na}$ are the ion-specific constant conductances, while $A$ and $B$ are the activation and inactivation gating variables, respectively, where $\alpha$ and $\beta$ are their respective constants. This means that, both $\bar{g}_{\rm K}$ and $\bar{g}_{\rm Na}$ are the intrinsic conductances. We will address the reasons why and how these conductances can be different from each other microscopically and unambiguously for a given ion channel (when we address points (I) and (II) listed above).\\
\textit{Example 2}: Within the Poisson-Nernst-Planck-Boltzmann (PNPB) formalism, one starts from a Poisson equation,
\begin {eqnarray}
&&\epsilon_0\nabla\cdot\big[\epsilon(\textbf{r})\nabla\varphi(\textbf{r})\big] = -\sum_i\rho_i, \label{eq:12}
\end {eqnarray}  
where $\rho$ is the charge density arising from the scalar potential, $\varphi(\textbf{r})$ such that $i$ counts the types of charge density (from electrons ($\rho_{\rm el}$), ions ($\rho_{\rm ion}$), and other external sources ($\rho_{\rm ext}$)), $\textbf{r}$ is the charge coordinate, $\epsilon_0$ is the permittivity of free space, and $\epsilon(\textbf{r})$ is the dielectric function. In continuum theoretical approaches~\cite{bastug}, $\epsilon(\textbf{r})$ is usually taken to be a constant ($\epsilon$), for example, $\epsilon_{\rm water} \approx 80$, $\epsilon_{\rm protein} \approx 2$ and $\epsilon_{\rm vacuum} \approx 1$. Thus, Eq.~(\ref{eq:12}) simplifies to       
\begin {eqnarray}
&&\epsilon_0\epsilon\big[\nabla^2\varphi(\textbf{r})\big] = -\sum_i\rho_i = \epsilon_0\epsilon x^2\varphi, \label{eq:13}
\end {eqnarray}  
\begin {eqnarray}
&&x^{-1} = \sqrt{\frac{\epsilon_0\epsilon k_{\rm B}T}{2z^2e^2n_0}}, \label{eq:14}
\end {eqnarray}  
where $x^{-1}$ is the Debye screening length, $T$ is the temperature in Kelvin, $z$ counts the number of charges, $k_{\rm B}$ and $n_0$ denote the Boltzmann constant and the charge-carrier number density (a constant), respectively. The term, $x^2\varphi$ actually originated from the Debye-H${\rm \ddot{u}}$ckel approach~\cite{debye}. In this PNPB approach, the current of each ion species is determined from the ion flux~\cite{cory}, 
\begin {eqnarray}
&&J = -D\bigg[\nabla n + \frac{zen}{k_{\rm B}T}\nabla\varphi\bigg]. \label{eq:15}
\end {eqnarray}  
Here, $D$ is the diffusion coefficient and $n$ is the charge carrier number density (not a constant). In this PNPB formalism however, we have two intrinsic constants, $\epsilon$ and $D$. Recall here that the reason why we call them as ``intrinsic constants'' is because they implicitly depend on some microscopic parameters associated to electronic wavefunctions or energy levels, and they are constants for a given system under certain conditions. Since $D$ refers exclusively to ions, we can treat it as a macroscopic constant as required such that all the microscopic electronic effects are allowed to be handled by $\epsilon$. In fact, we have provided the procedure to treat $\epsilon$ as a microscopic function in our earlier work~\cite{pla2}, for example, we have developed a phenomenological theory of dielectric function ($\epsilon$) within IET, which formally treats $\epsilon$ as a microscopic function that depends on the electronic energy levels~\cite{pla2}. In contrast, treating $\epsilon$ as an intrinsic constant as was done in the PNPB formalism makes it difficult to be used to address the points stated earlier in (I) and (II).\\
\textit{Example 3}: In Brownian dynamics simulations of individual ions, one often works with the Langevin-type equation~\cite{cory2}, 
\begin {eqnarray}
&&m_i\frac{{\rm d}v_i}{{\rm d}t} = -m_i\gamma_iv_i + F_{\rm random}(t) + z_ie\textbf{E}_i + F_{\rm range}^{\rm short}, \label{eq:16}
\end {eqnarray}  
where $m_i$ and $v_i$ are the mass and velocity of the $i^{\rm th}$ ion, respectively, $\gamma$ denotes the coefficient of friction, $\textbf{E}$ is the electric field experienced by an ion, $F_{\rm random}$ is the force acting on an ion due to random collisions, and $F^{\rm short}_{\rm range}$ represents the collection of some short-range forces. Apparently, $\gamma$ is the only parameter that can be associated implicitly to quantum physics, and therefore, $\gamma$ is an intrinsic constant. For example, $\gamma$, which defines the frictional force experienced by an ion can be shown to exist due to electron-electron and electron-ion Coulomb forces (both attractive and repulsive Coulomb forces). Microscopically, these Coulomb forces are the ones that give rise to a frictional force experienced by an ion. In any case, in the absence of a proper microscopic definition for $\gamma$, Eq~(\ref{eq:16}) cannot lead us to solve the problems listed in (I) and (II).\\ 
\textit{Example 4}: Simulations carried out with the molecular dynamics (MD) have got nothing to do with quantum mechanical method because MD method does not deal with wave functions nor any electronic Hamiltonians~\cite{ira}. For example, users will decide which atom in a given molecule is bonded to which atom, and the types of bond, and also the coordinates for these atoms in that molecule. 

Add to that, this molecular-mechanics method considers a molecule as composed of atoms with bonds that allow bending, stretching, torsion, and other important interactions---vdW, non-diagonal and electrostatic interactions. Non-diagonal interaction here means an interaction due to coupling of two different physical phenomena, for example the coupling of electronic and phononic parameters. In this case, the electron-phonon coupling cannot be decoupled because they are coupled non-adiabatically, which needs to be treated as a non-diagonal type interaction. Hence, MD method calculates the changes to the molecule's electronic energy from the above-stated interactions. If one incorporates some quantum mechanical calculations into MD by evaluating some of the interactions, then one obtains the hybrid MD/QM method. This MD/QM approach can in principle handle the problems of ion selectivity beyond the thermodynamical and kinetic approaches. However, this method necessarily involves the use of guessed functions and also parameters that are needed to be adjusted variationally~\cite{ira,ptp,dna}. Details are given in these Refs.~\cite{ira,ptp,dna}, and we will not reproduce them here. The point here is, even though MD/QM can reproduce the Eisenman sequences, in principle, but it cannot explain why and how such sequences can even exist at all (down to an electronic level) due to the existences of guessed wavefunctions and variationally adjustable parameters in MD/QM or in any \textit{ab-initio} QM calculations. 

\subsection{Mechanisms of ion selectivity}

From the above four examples, it should be clear by now why we have opted to use IET to study the ion selectivity in ion channels. Now, we have reached the stage where we can expose the physical properties or conditions that are responsible to controlling the ion selectivity in cation-favoring ion channels (specifically, KcsA, Na$_{\rm v}$Rh and the L-type Ca$_{\rm v}$). The generalized conditions (or the generalized criteria responsible for ion selection) are---\\
(\texttt{a1}) Hydration energy of a particular cation~\cite{mac,clay,kin}.\\
(\texttt{a2}) Concentration of the ``correct'' cations when the gates are opened is large~\cite{lock} such that there exists an electrochemical gradient across the cell membrane.\\
(\texttt{a3}) Ligand-cation or carbonyl oxygen-cation attractive interactions (both Coulomb and vdW types) in accordance with Eqs.~(\ref{eq:7}),~(\ref{eq:8}) and~(\ref{eq:10}).\\
(\texttt{a4}) Pore size (from the crystal structure) and the diameter of a cation~\cite{mac,clay,kin}.

Having listed the required conditions, we would like to inform you that the physical condition stated in (\texttt{a3}) is one of our claim made in this work. This claim will be proven in the following analyses, with experimental supports from Refs.~\cite{xu,hess,will}. The second claim made here is that the ion selectivity must satisfy at least one of the above-listed conditions ((\texttt{a1}) to (\texttt{a4})) or any combination of them, or all of them. It may be surprising to some researchers that we have excluded the well-studied ion selectivity condition,\\
(\texttt{b5})---the thermodynamic stability and the kinetic approaches. 

Here, (\texttt{b5}) also incorporates the unique crystal structure of the selectivity filter within a KcsA ion channel, because both the free-energy and kinetic-path calculations require the knowledge of this crystal structure. The reason for excluding (\texttt{b5}) is because the low free-energy binding sites and the large knock-on strength path~\cite{hod2,kuyu,kuyu2,roux2} are not directly relevant to ion selectivity. 

For example, the two quantities stated above are not directly relevant because they only control the ionic conductance and the permeation of cations through the selectivity filter (with maximum conductance), but they (the free-energy and the knock-on mechanism) do not ``decide'' which cations can or cannot enter the selectivity pore. You may want to recall the negated ejection-mechanism explained earlier, and the analyses on the L-type Ca$_{\rm v}$. This means that, the well-studied condition, (\texttt{b5}) is a completely independent condition responsible for indirect ion selection within the selectivity filter (after the cations enter the selectivity filter). Whereas, the conditions listed in (\texttt{a1}) to (\texttt{a4}) are directly related to ion selectivity before the cations could enter the selectivity pore. In other words, (\texttt{a1}) to (\texttt{a4}) will ensure that the probability of the correct ions to enter the selectivity pore is large. One should note here that the ``incorrect ions'' (after entering the pore) may block the channel pore due to their low permeability (small conductance) through the selectivity filter, giving rise to the importance of (\texttt{b5}). In the subsequent analyses, we will explain why and how (\texttt{a3}) is responsible for the blocking mechanism and the relative magnitudes of the hydration energy. However, too much focus on (\texttt{b5}) using (\texttt{a1}) and (\texttt{a4}) have led us to a situation where ion selectivity is falsely assumed to be an effect of ion conductance or permeation through the selectivity pore such that the ion selection processes are not generalizable across the cation-favoring voltage-gated ion channels. 

\subsection{Analysis I: KcsA}

We have given sufficient details on the KcsA ion channels much earlier in the introduction (you may want to recall them before reading on). It is strange that a KcsA ion channel (pore diameter, 3 to 4~\AA) almost exclusively select a larger K$^+$ ion (2.67~\AA) to enter and permeate through the selectivity pore more efficiently than the smaller Na$^+$ ions (1.9~\AA)~\cite{ney,ney2,mac,cri}. Other monovalent and divalent cations are also easily excluded~\cite{ney,ney2,mac,cri}. For your convenience, all sizes are given in \AA~ and they refer to diameters, not radii. In KcsA ion channels, all the listed conditions for ion selectivity ((\texttt{a1}) to (\texttt{a4}), and (\texttt{b5})) come into play. In particular, (\texttt{a1}) makes it difficult for a fully hydrated Na$^+$ ion to get rid of water molecules, compared to an equally hydrated K$^+$ ion. Here ``equally hydrated'' means that both cations are surrounded by equal number of water molecules. 

The dehydration energy for a Na$^+$ ion is larger than that of a K$^+$ ion~\cite{beza,hille}, which can also be confirmed with IET (using Eqs.~(\ref{eq:7}),~(\ref{eq:8}),~(\ref{eq:10}) and~(\ref{eq:3})). Since both Na$^+$ and K$^+$ are monovalent cations, we are not required to use Eq.~(\ref{eq:3}) to obtain their respective ionization energies, instead we can directly use the raw data reported in the databases~\cite{web,web2} (see Table~\ref{Table:I}) without any averaging where $\xi_{\rm Na^+}$ = 496~kJmol$^{-1}$ and $\xi_{\rm K^+}$ = 419~kJmol$^{-1}$. Therefore, $\xi_{\rm Na^+} > \xi_{\rm K^+}$ and this inequality means (using Eq.~(\ref{eq:7})) that a Na$^+$ ion can attract an electron donor (the oxygen ion from a water molecule) more strongly compared to a K$^+$, even though electrostatically both (Na$^+$ and K$^+$) are 1+. This means that, a Na$^+$ ion (compared to a K$^+$ ion) is strongly bounded to a water molecule, which means the hydrated Na$^+$ ions require a larger dehydration energy. What we did above was to exploit (\texttt{a3}) in order to prove (\texttt{a1}) as a valid condition that assists a KcsA ion channel to select K$^+$. Using the same equation (Eq.~(\ref{eq:7})), we can easily show that all the monovalent ions with large $\xi$ (larger than $\xi_{\rm K^+}$) can be excluded due to (\texttt{a1}). On the other hand, all the divalent cations are also automatically excluded due to (\texttt{a1}) because the dehydration energy is always larger for divalent cations, compared to monovalent cations (as a result of stronger Coulomb-vdW interaction). 

Here we provide the proof (using IET) for the existence of the dehydration energy. The Coulomb-vdW attraction between a cation (Na$^+$ or K$^+$) and a oxygen from a water molecule is stronger than that of the Coulomb-vdW attractive strength between a cation (Na$^+$ or K$^+$) and a oxygen from a carbonyl group. The formal proof for this statement is available in Refs.~\cite{pccp,dna}, which can be determined from Eq.~(\ref{eq:7}). Briefly, the oxygen from a water molecule ($^{\delta-}$OH$_2$) is more negatively charged compared to the oxygen from a carbonyl group ($^{\delta'-}$OC(RH)) such that $\delta- > \delta'-$. The reason for this inequality is straightforward from IET where the oxygen from a water molecule is negatively charged as a result of the two electrons contributed by two hydrogen atoms. The averaged $\xi_{\rm H^+}$ for these two electrons is 1312~kJmol$^{-1}$ (see Table~\ref{Table:I}), which is smaller than the two electrons contributed by a carbon atom ($\xi_{\rm C^{4+}}$ = 3371~kJmol$^{-1}$) to a carbonyl oxygen. Therefore, the oxygen from a carbonyl group is less negatively charged ($\delta'-$) compared to the oxygen ($\delta-$) from a water molecule. This means that, the cations prefer to be surrounded by water molecules, compared to carbonyl oxygens, and therefore, one needs to dehydrate these cations so that they (the cations) can enter the selectivity pore. 

The valence state, namely, 1+ (H$^+$) here does not imply an electron has been completely removed from the atomic hydrogen, instead it means that this particular electron has been polarized or excited to a finite distance $r$, within the water molecule. If the valence state is 4+ (C$^{4+}$), then there are four polarized or excited electrons at distances $r_1$, $r_2$, $r_3$ and $r_4$, within the carbonyl group, and so forth. Note here that for a carbon atom, there are four electrons involved, two are polarized toward the oxygen atom, while the other two electrons are respectively polarized toward another atomic carbon and a hydrogen atom. This explains why we used $\xi_{\rm C^{4+}}$, and we cannot distinguish the electrons any further. 

Subsequently, it is also straightforward to verify (\texttt{a2}) as another valid condition for ion selectivity in cation-favoring voltage-gated ion channels. In particular, it is because of (\texttt{a2}) that one obtains a large number of K$^+$ ions (of the order of one million ions per second or higher) permeating through the KcsA selectivity filter. We note here that (\texttt{a2}) is true regardless whether a large concentration of K$^+$ is or is not required (as a stimulant) to activate the voltage-sensitive KcsA gates to open (or to remain open). 

Although (\texttt{a3}) has been used indirectly to prove (\texttt{a1}), but we did not show why and how (\texttt{a3}) is valid as an independent condition, required for K$^+$ ion selection in a KcsA ion channel. For example, to prove (\texttt{a1}), we used Eq.~(\ref{eq:7}) to justify the attraction between the cations (K$^+$ and Na$^+$) and the oxygen from a water molecule, not the oxygen from the carbonyl groups. We will show (\texttt{a3}) is an independent criterion when we discuss Na$_{\rm v}$Rh and Ca$_{\rm v}$ later because we cannot prove it to be independent using the KcsA channels. The attractive interaction between a cation (Na$^+$ or K$^+$) and the carbonyl group oxygens (within the selectivity filter) determines the accumulation of cations with large $\xi$ and charge ($Z+$) near the selectivity pore (before entering the pore). It is straightforward from Eq.~(\ref{eq:7}) to deduce that the cations with large $\xi$ and $+Ze$ are maximally attracted towards the negatively charged carbonyl oxygens (from the narrow selectivity pore). This implies that Na$^+$ ions (compared to K$^+$) are the ones that should be accumulating near the selectivity filter (before entering). Fortunately, this accumulation does not mean that the KcsA ion channel select Na$^+$ ions over K$^+$ ions (see below).

Thanks to $\xi_{\rm Na^+} > \xi_{\rm K^+}$, there are more water molecules surrounding a Na$^+$ ion than a K$^+$ ion because of a stronger Coulomb-vdW attraction between a Na$^+$ cation and an oxygen (from a water molecule). This means that, a Na$^+$ ion is screened by a large number of water molecules compared to a K$^+$ ion, and therefore both cations have identical Coulomb-vdW attractive strengths between a cation (Na$^+$ or K$^+$) and a carbonyl oxygen (not the oxygen from a water molecule). Therefore, Na$^+$ ions may not have the tendency to accumulate near the narrow pore because the Na$^+$ ions have been more strongly screened (compared to K$^+$ ions) by a large number of water molecules, giving rise to an identical Coulomb-vdW attractive strengths between a cation (Na$^+$ or K$^+$) and a carbonyl oxygen, even though $\xi_{\rm Na^+} > \xi_{\rm K^+}$.

We now counter the above arguments that claim Na$^+$ ions have been more strongly screened compared to K$^+$ ions leading to identical strength in the Coulomb-vdW attraction between both cations and the carbonyl oxygens (from the selectivity pore). Our counter argument reads---both cations are equally screened such that $\xi_{\rm Na^+} > \xi_{\rm K^+}$ is still valid, giving rise to a stronger Coulomb-vdW attraction between a Na$^+$ (compared to a K$^+$) and a carbonyl oxygen. The counter-argument will be proven to be correct when we prove the Eisenman sequences and the blocking of Na$^+$ current by Cd$^{2+}$ and Ca$^{2+}$ ions in the following sections. Anyway, the inner-layer water molecules are the ones responsible for the hydration of cations, and these (inner-layer) water molecules give rise to equal screening for both Na$^+$ and K$^+$. The water molecules that form the inner-layer (that have surrounded a cation) remain the same all the time due to large dehydration energy (unless the cations are dehydrated). On the other hand, the outer-layer water molecules act only as carriers such that they can always be removed or switched with other mobile water molecules in the vicinity with a much smaller dehydration energy. This means that, $\xi_{\rm Na^+} > \xi_{\rm K^+}$ is indeed active and is responsible for the higher tendency of the hydrated Na$^+$ ions (compared to the hydrated K$^+$) to accumulate at the entrance of the selectivity pore. However, the hydrated Na$^+$ ions accumulated near the selectivity pore do not block the pore because they cannot enter this narrow pore, without being at least partially dehydrated (removal of the inner-layer water molecules). Subsequently, due to large dehydration energy and mobile water molecules, their (Na$^+$ ions) accumulation does not block the hydrated K$^+$ from reaching the selectivity pore. Apart from that, in the presence of (\texttt{a2}), disturbances from Na$^+$ ions can be negligible. 

The above analyses lead us to conclude that both (\texttt{a1}) and (\texttt{a2}) have given the advantage to K$^+$ ions to enter the KcsA selectivity pore, despite the fact that more Na$^+$ ions (compared to K$^+$) can reach the narrow pore due to a stronger Coulomb-vdW attraction between a Na$^+$ and a carbonyl oxygen. However, the influence of (\texttt{a3}) is indirect and it still favors K$^+$ ions by means of a weaker Coulomb-vdW attraction, K$^+$$---$OH$_2$, which require a lower dehydration energy compared to Na$^+$$---$OH$_2$ (recall the analyses on (\texttt{a1})). Using the same arguments explained above, we can exclude all other monovalent (except H$^+$ ions) and divalent cations from passing through the KcsA ion channels. Due to their small size, hydrated H$^+$ ions can pass through the KcsA, Na$_{\rm v}$Ab and the L-type Ca$_{\rm v}$ ion channels without the need to dehydrate itself (the dehydration energy for H$^+$ is large compared to other monovalent cations listed in Table~\ref{Table:I} because $\xi_{\rm H^+}$ is relatively large).   

The two parameters (ionic and pore sizes) in (\texttt{a4}) are related to the ability of residues (negatively charged) in the selectivity filter to stretch closer to a cation (due to Coulomb-vdW attraction), but not away from the cation. This means that, the selectivity filters are always flexible (to some extent) in the direction towards a cation. The ``flexible residue'' above means flexible bonds due to polarizable bonding electrons, and they are not related to classical stretching in any way. This means that the ions with sizes equal or larger than the pore size cannot enter the selectivity pore. In other words, the pore size does not get bigger to accommodate a larger ion. These same arguments also apply for the hydrated ions, provided that the dehydration energy is large enough.

So far, we have evaluated the correctness of all the physical conditions ((\texttt{a1}) to (\texttt{a4})), which are in play before the cations could enter the narrow selectivity pore. Once the cations (the correct ones) are in the pore, (\texttt{b5}) (a level-two condition) becomes active and determines the conducting path with maximum conductance. We stress here that the level-one conditions ((\texttt{a1}) to (\texttt{a4})) do not exclusively select the correct cations to enter the filter, instead they give rise to the highest probability for the correct ions to enter the selectivity pore. Consequently, there are small chances for the incorrect cations (namely, Na$^+$ ions) to enter the KcsA selectivity pore. In this case, (\texttt{b5}) will ensure Na$^+$ ions do not conduct as easily as K$^+$ ions, and therefore, the current due to Na$^+$ ions ($I_{\rm Na^+}$) is always smaller than $I_{\rm K^+}$ in the KcsA ion channels. The reasons for the high conductance for K$^+$ ions (compared to Na$^+$ ions) have been exposed in the previous studies as due to the knock-on assisted permeation with the lowest free-energy pathway~\cite{hod2,kuyu,kuyu2,roux2}. 

We can also invoke (\texttt{a3}) to explain why $I_{\rm Na^+} < I_{\rm K^+}$ is true. We first substitute the inequality $\xi_{\rm Na^+} > \xi_{\rm K^+}$ into Eq.~(\ref{eq:8}) and subsequently, insert Eq.~(\ref{eq:8}) and~(\ref{eq:10}) into Eq.~(\ref{eq:7}) to obtain the strength of Coulomb-vdW attraction between a cation (Na$^+$ or K$^+$) and a carbonyl oxygen. We find that $\tilde{V}'_{\rm Waals}(\xi)$ (see Eq.~(\ref{eq:7})) for Na$^+$ ions is larger than for K$^+$ ions, and therefore, a Na$^+$ ion is strongly bounded to a carbonyl oxygen (compared to a K$^+$), which then leads us to $I_{\rm Na^+} < I_{\rm K^+}$. Apparently, strongly bounded Na$^+$ ions can block the KcsA selectivity pore to some extent, depending on the strength of $\tilde{V}'_{\rm Waals}(\xi,\rm Na^+)$ compared to $\tilde{V}'_{\rm Waals}(\xi,\rm K^+)$ where $\tilde{V}'_{\rm Waals}(\xi,\rm Na^+) > \tilde{V}'_{\rm Waals}(\xi,\rm K^+)$. This blocking mechanism will be further explained when we discuss the Eisenman sequences for the L-type Ca$_{\rm v}$ ion channels. We are now ready to tackle the Na$_{\rm v}$Rh ion channels, which should be easier because they too, obey all the notions introduced for the KcsA ion channels.           

\subsection{Analysis II: Na$_{\rm v}$Rh}

We have understood the microscopic mechanisms that come to play via the conditions, (\texttt{a1}) to (\texttt{a4}) and (\texttt{b5}), which are responsible for the K$^+$-ion selection in a KcsA channel, and how the KcsA selectivity pore exclude the Na$^+$ ions. We can now move on to explore the cation selection in a Na$_{\rm v}$Rh ion channel. We exclusively choose the Na$_{\rm v}$Rh ion channels as reported by Zhang et al.~\cite{xu} because they have measured the conductance for this channel in the presence of different ions---Na$^+$, K$^+$, Cs$^+$, Cd$^{2+}$, Ba$^{2+}$ and Ca$^{2+}$. The narrowest pore~\cite{xu} within a Na$_{\rm v}$Rh selectivity filter is around 1.84~\AA~ to 2.12~\AA, which is much less than the diameter of a dehydrated K$^+$ ion (2.67~\AA). Using (\texttt{a4}), we can readily deduce that K$^+$ ions cannot enter and permeate through a Na$_{\rm v}$Rh selectivity pore, which is indeed the case here (see Fig.1b and Fig.2f in Ref.~\cite{xu}). Apart from K$^+$ ions, (\texttt{a4}) also excludes other larger ions (diameter larger than 2.12~\AA), namely, Rb$^+$, Cs$^+$, Sr$^{2+}$ and Ba$^{2+}$ (see Table~\ref{Table:I}).  

On the other hand, the conditions (\texttt{a3}) and (\texttt{a4}) have made sure that the partially or fully dehydrated Na$^+$ ions are the ones that can permeate through a Na$_{\rm v}$Rh ion channel with the highest conductance because (i) $\xi_{\rm Na^+}$(496~kJmol$^{-1}$) $<$ $\xi_{\rm Ca^{2+}}$(868~kJmol$^{-1}$) $<$ $\xi_{\rm Cd^{2+}}$(1250~kJmol$^{-1}$) and (ii) \O$_{\rm Na^+}$(1.9~\AA) $<$ \O$_{\rm Cd^+}$(1.94~\AA) $<$ \O$_{\rm Ca^{2+}}$(1.98~\AA) $<$ \O$^{\rm Na_vRh}_{\rm pore}$(2.12~\AA) $<$ \O$_{\rm K^+}$(2.67~\AA) $<$ \O$_{\rm Ba^{2+}}$(2.7~\AA) $<$ \O$_{\rm Cs^+}$(3.38~\AA) where \O~ denotes the diameter. The selectivity pore of a Na$_{\rm v}$Rh channel cannot prevent a hydrated H$^+$ from entering, but we will not consider H$^+$ ions any further due to lack of available experimental data. The condition, (\texttt{a2}) implies that there is a large number of Na$^+$ ions (relative to other cations) nearby a cell membrane, which are ready to permeate through the Na$_{\rm v}$Rh channels, which further reinforces the magnitude of Na$^+$-ion current ($I_{\rm Na^+}$). Here, the Coulomb-vdW attraction between a Na$^+$ ion and a carbonyl oxygen (from the Na$_{\rm v}$Ab selectivity pore) is larger than that of a K$^+$ ion. Hence, (\texttt{a3}) promotes the accumulation Na$^+$ ions in the vicinity of a Na$_{\rm v}$Rh selectivity filter, which further justifies (\texttt{a2}). However, the inequality in (i) above unequivocally proves that Cd$^{2+}$ and Ca$^{2+}$ ions accumulation rate at the entrance of the selectivity pore is much higher than that of Na$^+$ ions. Apart from that, the dehydrated Cd$^{2+}$ and Ca$^{2+}$ can enter the selectivity filter because the ionic sizes for Na$^+$(1.9~\AA), Cd$^{2+}$(1.94~\AA) and Ca$^{2+}$(1.98~\AA) are close to each other. After entering, Cd$^{2+}$ ions can block $I_{\rm Na^{+}}$ more effectively compared to Ca$^{2+}$ ions because $\xi_{\rm Ca^{2+}}$ $<$ $\xi_{\rm Cd^{2+}}$. This means that the attractive interaction between a Cd$^{2+}$ ion and a carbonyl oxygen is the largest compared to a Na$^+$ or a Ca$^{2+}$ ion, which have been shown experimentally to be true as depicted in Fig.2f in Ref.~\cite{xu}. 

The condition related to the dehydration energy, (\texttt{a1}) can be verified with (\texttt{a3}), somewhat identical to the previously discussed KcsA ion channels. For example, from this inequality, $\xi_{\rm Na^+}$(496~kJmol$^{-1}$) $<$ $\xi_{\rm Ca^{2+}}$(868~kJmol$^{-1}$) $<$ $\xi_{\rm Cd^{2+}}$(1250~kJmol$^{-1}$), we can easily deduce that both Ca$^{2+}$ and Cd$^{2+}$ ions need higher dehydration energies compared to Na$^+$ ions. Therefore, (\texttt{a3}) suppresses the probability for Ca$^{2+}$ and Cd$^{2+}$ ions to enter the selectivity pore of a Na$_{\rm v}$Rh channel even though both Cd$^{2+}$ and Ca$^{2+}$ ions can accumulate (again due to (\texttt{a3})) at the entrance of a selectivity filter faster than Na$^+$ ions. The other ions that have lower ionization energies (K$^+$, Rb$^+$ and Cs$^+$) compared to $\xi_{\rm Na^+}$ also have lower probabilities to enter the selectivity filter because of their large sizes, thus (\texttt{a4}) excludes these low ionization-energy ions from entering and permeating through the Na$_{\rm v}$Rh channels. Hence, we have made clear here why and how these level-one conditions ((\texttt{a1}) to (\texttt{a4})) nicely play their parts to make sure that only the ``correct'' ions have the maximum probability to enter and permeate with maximum conductance. However, we have excluded (\texttt{b5}) from consideration because (\texttt{a1}) to (\texttt{a4}) are sufficient to understand the mechanism of ion selection in a Na$_{\rm v}$Rh ion channel. Of course (\texttt{b5}) is important if we decide to evaluate the ionic conductance curves, which is not our objective here because our intention is not to reproduce the conductance curves, which have been done by others (see the four examples given in the previous section). 

\subsection{Analysis III: L-type Ca$_{\rm v}$} 

Evaluating the permeation of Ca$^{2+}$ ions through a L-type Ca$_{\rm v}$ channel also require us to invoke the generalized conditions, (\texttt{a1}) to (\texttt{a4}) without the need to know the independent mechanism, (\texttt{b5}) that is associated to ion permeation within the selectivity pore. What matters here, is that we need to prove the correctness of the Eisenman sequences~\cite{eisen} measured in the L-type Ca$_{\rm v}$ ion channels~\cite{hess,will} in the presence of different ions such as Na$^+$, K$^+$, Cs$^+$, Cd$^{2+}$, Ba$^{2+}$ and Ca$^{2+}$. In L-type Ca$_{\rm v}$ ion channels, we do not have the data on the pore diameter, which means that we cannot invoke (\texttt{a4}) and therefore, our focus is to study and justify the Eisenman sequences alone using IET and the relevant level-one conditions ((\texttt{a1}) to (\texttt{a3})). Thus far, we have learned that large ionization-energy ions can accumulate near the entrance of a selectivity pore, faster than the small ionization-energy ions. But the dehydration energies for these large $\xi$ ions are also larger, which gives rise to competing effects that influence the probability for these large $\xi$ ions to enter the selectivity filter. In other words, if these large $\xi$ ions can enter the selectivity pore (at least after partial dehydration), then they will block the permeation of the correct ions through the selectivity pore. Here, the correct ion is Ca$^{2+}$ that has a lower ionization energy ($\xi_{\rm Ca^{2+}}$).       

The reversal potential, $E_{\rm rev}$ measurements for both monovalent and divalent cations have been used to determine the Eisenman sequence~\cite{hess,will}, namely, Ca$^{2+}$ $>$ Ba$^{2+}$ $>$ Li$^{+}$ $>$ Na$^{+}$ $>$ K$^{+}$ $>$ Cs$^{+}$. This sequence implies that Ca$^{2+}$ ions have the lowest permeation rate, while a Cs$^+$ ion permeates the L-type Ca$_{\rm v}$ selectivity filter with the highest permeation rate~\cite{hess,will}. The reason for this is due to large binding energy for a Ca$^{2+}$ ion in the selectivity filter, and this binding energy reduces systematically from Ca$^{2+}$ to Cs$^{+}$, giving rise to the above Eisenman sequence~\cite{hess,will}. Apparently, one can use the MD/QM method to calculate the above binding energies and reproduce the said sequence. However, we need to dig deep to understand why the binding energies of these cations has to follow the Eisenman sequence. Meaning, we will need to answer this question---what is the physical mechanism that is responsible to produce such a well-defined binding-energy sequence (down to an electronic level)? Obviously, this question is beyond the reach of any \textit{ab-initio} QM method~\cite{ira}.

Before answering the above question, which is actually very straightforward within IET, we should also be aware here that one can obtain the conductance sequence from the above reversal-potential Eisenman sequence, which is in the reverse order (as it should be), Ca$^{2+}$ $<$ Ba$^{2+}$ $<$ Li$^{+}$ $<$ Na$^{+}$ $<$ K$^{+}$ $<$ Cs$^{+}$. This conductance sequence means that a Ca$^{2+}$ ion has the lowest conductance because it also has the largest binding energy, whereas, a Cs$^+$ ion (with the smallest binding energy) can permeate through a L-type Ca$_{\rm v}$ selectivity filter with the fastest permeation rate~\cite{hess,will}. Using Eq.~(\ref{eq:3}), we can calculate the ionization energies for all the cations that appear in the reversal potential sequence, and is given by Ca$^{2+}$(868~kJmol$^{-1}$) $>$ Ba$^{2+}$(734~kJmol$^{-1}$) $>$ Li$^{+}$(520~kJmol$^{-1}$) $>$ Na$^{+}$(496~kJmol$^{-1}$) $>$ K$^{+}$(419~kJmol$^{-1}$) $>$ Cs$^{+}$(376~kJmol$^{-1}$), which is nothing but the original experimentally-measured Eisenman sequence. Subsequently, we can substitute this sequence into Eq.~(\ref{eq:8}), and then use it together with Eq.~(\ref{eq:10}) to obtain Eq.~(\ref{eq:7}). In doing so, we can immediately observe (from Eq.~(\ref{eq:7})) that the maximum Coulomb-vdW attraction is between a Ca$^{2+}$ and a carboxyl oxygen (from the selectivity filter), while the minimum attraction is between a Cs$^+$ and the same carboxyl oxygen. In other the words, the so-called binding energies used in Refs.~\cite{hess,will} to explain the Eisenman sequence exist due to this (Coulomb-vdW) attraction, without the formation of any chemical bonds between any of these ions and a carboxyl oxygen. In fact, this attraction is the generalized hydrogen bond~\cite{pccp,dna}. Apart from that, Ba$^{2+}$, Li$^{+}$, Na$^{+}$, K$^{+}$ and Cs$^{+}$ cannot block the permeation of Ca$^{2+}$, instead we require ions with $\xi > \xi_{\rm Ca^{2+}}$ such as Mg$^{2+}$, Co$^{2+}$ and Cd$^{2+}$ (see Table~\ref{Table:I}) to block Ca$^{2+}$, provided that these large ionization-energy ions satisfy condition (\texttt{a1}). 

\section{conclusions}

We used logic to list all the conditions that are necessary to generalize the mechanisms of ion selectivity in cation-favoring voltage-gated ion channels of different types (in the open configuration), namely, KcsA, Na$_{\rm v}$Rh and the L-type Ca$_{\rm v}$. The conditions are further broken into two levels---the first level consists of the conditions, (\texttt{a1}) to (\texttt{a4}), which are valid before the cations enter the selectivity pore, while the second level consists of only one condition, (\texttt{b5}), which captures the ion selectivity indirectly within the selectivity filter (when the cations are within the selectivity pore). The reason why (\texttt{b5}) is only indirectly responsible for ion selectivity is because the incorrect cations are never ejected out of the pore (means, out of the cell, not into the cell), once they are found to be within the filter. The level-one conditions ((\texttt{a1}) to (\texttt{a4})) are related to (i) dehydration energy, (ii) concentration of the correct ion, (iii) Coulomb-vdW attraction, and (iv) pore and ionic sizes, respectively. Whereas, the level-two condition, (\texttt{b5}) is associated to the well known criterion---the thermodynamic stability and the knock-on permeation mechanism. Here, (\texttt{b5}) determines the the ionic conductance within the selectivity pore, while (\texttt{a1}) to (\texttt{a4}) ``decide'' which cation can enter the selectivity pore with the highest probability.

Next, we used the renormalized screened Coulomb and the stronger van der Waals attractive interactions within the ionization energy theory (IET) and the energy-level spacing renormalization group method to show that the mechanisms responsible for ion selectivity can be generalized using the above conditions ((\texttt{a1}) to (\texttt{a4}) and (\texttt{b5})). This means that, the logic used to generalize and state the level-one conditions ((\texttt{a1}) to (\texttt{a4})) has been unambiguously verified to be correct using the proper physico-chemical notions (within IET) to explain why and how each condition plays its crucial role in selecting the correct ion to enter the selectivity pore ((\texttt{a1}) to (\texttt{a4})) and permeate with maximum conductance ((\texttt{b5})). Hence, we have shown why and how the cation-favoring voltage-gated ion channels make use of the laws of quantum physics to select the correct ions to permeate through the selectivity filter. 

We also have shown why and how the condition (\texttt{a3}) can be generalized such that it can be used to justify the correctness of (\texttt{a1}) and (\texttt{b5}), unequivocally. Subsequently, we proved the logical and theoretical validity of (\texttt{a3}) with the available experimental observations. For example, (\texttt{a3}) is shown to be valid beyond any reasonable doubt because it correctly predicts that (i) Cd$^{2+}$ ions can block the conductance of Na$^+$ ions (in a Na$_{\rm v}$Rh ion channel) more effectively than the Ca$^{2+}$ ions possibly could, and (ii) reproduces the experimentally determined Eisenman sequences perfectly in the L-type Ca$_{\rm v}$ ion channels. In summary, we have derived a comprehensive theory that consists of well-defined conditions ((\texttt{a1}) to (\texttt{a4}) and (\texttt{b5})), in which (\texttt{a1}), (\texttt{a3}) and (\texttt{b5}) are microscopically related to atomic energy levels. Using these conditions, we have evaluated the three most well-studied cation-favoring voltage-gated ion channels unambiguously, namely, the KcsA, Na$_{\rm v}$Rh and the L-type Ca$_{\rm v}$ channels. Our theory is based on the energy-level spacing renormalization group method and it has not lead us to any self-contradiction or any violation with the experimental results. Most importantly, we did not invoke any patch, in any way, to enforce unequivocal agreement with the experimental observations.         

\section*{Acknowledgments}

This work was supported by Sebastiammal Innasimuthu, Arulsamy Innasimuthu, Arokia Das Anthony, Amelia Das Anthony, Malcolm Anandraj and Kingston Kisshenraj. This work was also supported by Kurunathan Ratnavelu through the University of Malaya research grant RG089/10AFR (from 18 June 2012 to 17 September 2012). I am grateful to Naresh Kumar Mani for Ref.~\cite{kandel}.

\end{document}